\documentclass[smallcondensed,natbib]{svjour3}     %
\smartqed  %
\usepackage{graphicx}
\graphicspath{{./Figures/}}

\usepackage{color}

\usepackage{amssymb}
\usepackage{amsfonts}
\usepackage{enumerate} 
\usepackage{array}
\usepackage{amsmath}
\usepackage[colorlinks=true,linkcolor=blue,citecolor=blue,urlcolor=blue]{hyperref}
\renewcommand{\epsilon}{\varepsilon}

\newcommand{\eg}{\emph{e.g.}}
\newcommand{\nnb}{\nonumber\\}

\newcommand{\Gr}{\mathcal{G}}
\renewcommand{\H}{\mathcal{H}}
\newcommand{\K}{\mathcal{K}}
\newcommand{\Z}{\mathbb{Z}}

\renewcommand{\i}{\iota}

\newcommand{\EXP}{\mathbf{e}} %

\newcommand{\lapc}[3]{b^{(#2)}_{#1}(#3)}

\renewcommand{\vec}[1]{\mathbf{#1}}

\newcommand{\AMD}{\ensuremath{C}}

\newcommand{\AM}{\ensuremath{G}}
\newcommand{\DG}{\ensuremath{\Delta G}}

\newcommand{\scalefactor}{\ensuremath{\Gamma}}

\newcommand{\hk}{h_1^{(\mathbf{k})}}
\renewcommand{\d}{\mathrm{d}}
\newcommand{\smallvar}[1]{\d#1}

\newcommand{\perrat}[1]{\ensuremath{\nu_{#1}}}

\newcommand{\dpart}[2]{\frac{\partial #1}{\partial #2}}

\journalname{Celestial Mechanics and Dynamical Astronomy}
\begin{document}

\title{An integrable model for first-order three-planet mean motion resonances}
\subtitle{}

\titlerunning{First-order three-planet MMR}        %

\author{Antoine C. Petit}

\institute{Antoine C. Petit \at
              Lund Observatory, Department of Astronomy and Theoretical Physics, Lund University, Box 43, 22100 Lund, Sweden   \\
              \email{antoine.petit@astro.lu.se}           %
}

\date{Received: date / Accepted: date}

\maketitle

\begin{abstract}
Recent works on three-planet mean motion resonances (MMRs) have highlighted their importance for understanding the details of the dynamics of planet formation and evolution.
While the dynamics of two-planet MMRs are well understood and approximately described by a one degree of freedom Hamiltonian, little is known of the exact dynamics of three-bodies resonances besides the cases of zeroth-order MMRs or when one of the body is a test particle.
In this work, I propose the first general integrable model for first-order three-planet mean motion resonances. I show that one can generalize the strategy proposed in the two-planet case to obtain a one degree of freedom Hamiltonian.
The dynamics of these resonances are governed by the second fundamental model of resonance.
The model is valid for any mass ratio between the planets and for every first-order resonance.
I show the agreement of the analytical model with numerical simulations.
As examples of application I show how this model could improve our understanding of the capture into MMRs as well as their role on the stability of planetary systems.

\end{abstract}

\section{Introduction}
\label{sec:intro}

Mean motion resonances (MMRs) are one of the keys of the dynamics of exoplanets.
Indeed, during planet formation, the disk-planet interactions \citep{Terquem2007} lead to the radial migration of the planet and eventually to capture into two-planet MMR \citep[\emph{e.g.}][]{Cresswell2008,Izidoro2017}.
However, comparisons to the observations suggest that at least 95\% of the resonant chains should break \citep{Izidoro2019}.
As a result, two-planet MMRs have been the subject of many researches over the past decade to understand the capture mechanism \citep{Ogihara2013,Batygin2015}, the breaking of the chains \citep{Matsumoto2012,Pichierri2018,Pichierri2020}, how they evolve in presence of tides \citep{Delisle2014a,Millholland2019a}, or their contribution to the instability of planetary systems \citep{Deck2013,Petit2017,Hadden2018}.

The dynamics of two-planet MMRs are now well understood. The dynamics of first-order MMRs are given by the so-called second fundamental model of resonance \citep{Henrard1983}.
The resonant dynamics of two massive planets on eccentric orbits are reduced to this one degree of freedom integrable Hamiltonian thanks to a sequence of transformations first introduced by \citet{Sessin1984} and \citet{Henrard1986}.
A generalization of this model for two-planet MMRs of arbitrary order has been proposed by \citet{Hadden2019}.

On the other hand, the study of three-bodies resonances has long been restricted to the asteroids dynamics in the Solar System \citep{Nesvorny1998,Cachucho2010}.
The question has gained a renewed interest thanks to the discovery of close-in exoplanet systems trapped into zeroth-order order resonances such as Trappist-1 \citep{Gillon2017,Agol2021} or more recently TOI-178 \citep{Leleu2021}.
Three-planet resonances are also thought to be the main driver of the early instability of tightly packed systems \citep{Quillen2011,Petit2020a}.

The thorough numerical studies by \citet{Guzzo2005} or \citet{Charalambous2018} has shown the key role of three-planet resonances in shaping the phase space.
Particularly, \citet{Charalambous2018} highlighted their role during resonance capture, where first-order three-planet MMRs act as a guide towards two-planet MMR during multiple planet migration.
Semi-analytical studies \citep{Gallardo2016} have also been performed and given a qualitative understanding of the three-planet resonance dynamics.

A general integrable model for zeroth-order three-planet MMRs is directly obtained from a second-order averaging since no transformation of the coordinates is needed besides the introduction of the resonant angle \citep[see][]{Petit2020a}.
However, this result is not immediately generalized to higher order MMRs.
The question of the existence of an integrable model for first-order three-planet MMRs is thus of theoretical interest.
Besides that, such a model could help better understand the capture into these resonances as well as their role into the instability of tightly packed systems with eccentric orbits.
To this day, there is no direct evidence of exoplanets trapped into first-order three-planet MMR.
Yet it seems that they could help stabilize some systems as suggested for Kepler-11 \citep{Migaszewski2012}.

In section~\ref{sec:network}, I introduce the different networks of first-order three-planet resonances and discuss their relative importance with respect to other two-planet and three-planet MMRs.
I show in section~\ref{sec:model} how to obtain a one degree of freedom Hamiltonian for the dynamics of an isolated first-order three-planet MMR, adapting the results from the two-planet case.
I then analyse the properties of the resonance thanks to this integrable model (section \ref{sec:width}).
Finally, I present in section~\ref{sec:applications} some examples of application of the model to the problem of stability of eccentric tightly packed systems and the capture into these resonances, while leaving the detailed studies for future work.

\section{Resonance networks}
\label{sec:network}

First-order three-planet resonances occur when the mean motions verify the relationship 
\begin{equation}
	k_1n_1+k_2n_2+k_3n_3 = 0, \label{eq:res}
\end{equation}
where $k_1+k_2+k_3 = 1$ and $n_j$ are the Keplerian mean motions. 
The resonance equation defines a plane in the frequency space $(n_1,n_2,n_3)$.
Because the gravitational interactions are scale invariant, we can restrict ourselves to a two-dimensional plane corresponding to the period ratios \perrat{12} and \perrat{23} where 
\begin{equation}
\perrat{ij} = \frac{P_i}{P_j}= \frac{n_j}{n_i}.
\label{eq:perrat}
\end{equation}
The resonance relation can be rewritten to remove $k_2$ and use period ratios in place of the mean motions.
In the general case, one obtains
\begin{equation}
	\perrat{23} = 1-\frac{1}{k_3} -\frac{k_1}{k_3}\left(\perrat{12}^{-1}-1\right),
	\label{eq:firstorderrel}
\end{equation}
which is the equation of the locus of the unperturbed resonance defined by $k_1$ and $k_3$.
One can remark that except for $-k_3^{-1}$ on the right-hand side, it is similar to the zeroth-order resonance relationship used in \citep{Petit2020a}.
However, unlike the zeroth-order case, there are three different families of resonances depending on the signs of $k_1$ and $k_3$.
Noting $p=|k_1|$ and $q=|k_3|$, we have
\begin{align}
	\perrat{23} &= 1-\frac{1}{q} -\frac{p}{q}\left(\perrat{12}^{-1}-1\right), &\quad k_1>0, k_3>0,\label{eq:blue}\\
	\perrat{23} &= 1+\frac{1}{q} -\frac{p}{q}\left(\perrat{12}^{-1}-1\right), &\quad k_1<0, k_3<0,\label{eq:red}\\
	\perrat{23} &= 1+\frac{1}{q} +\frac{p}{q}\left(\perrat{12}^{-1}-1\right), &\quad k_1>0, k_3<0\label{eq:green}.
\end{align}
\begin{figure}\centering
	\includegraphics[width=0.8\linewidth]{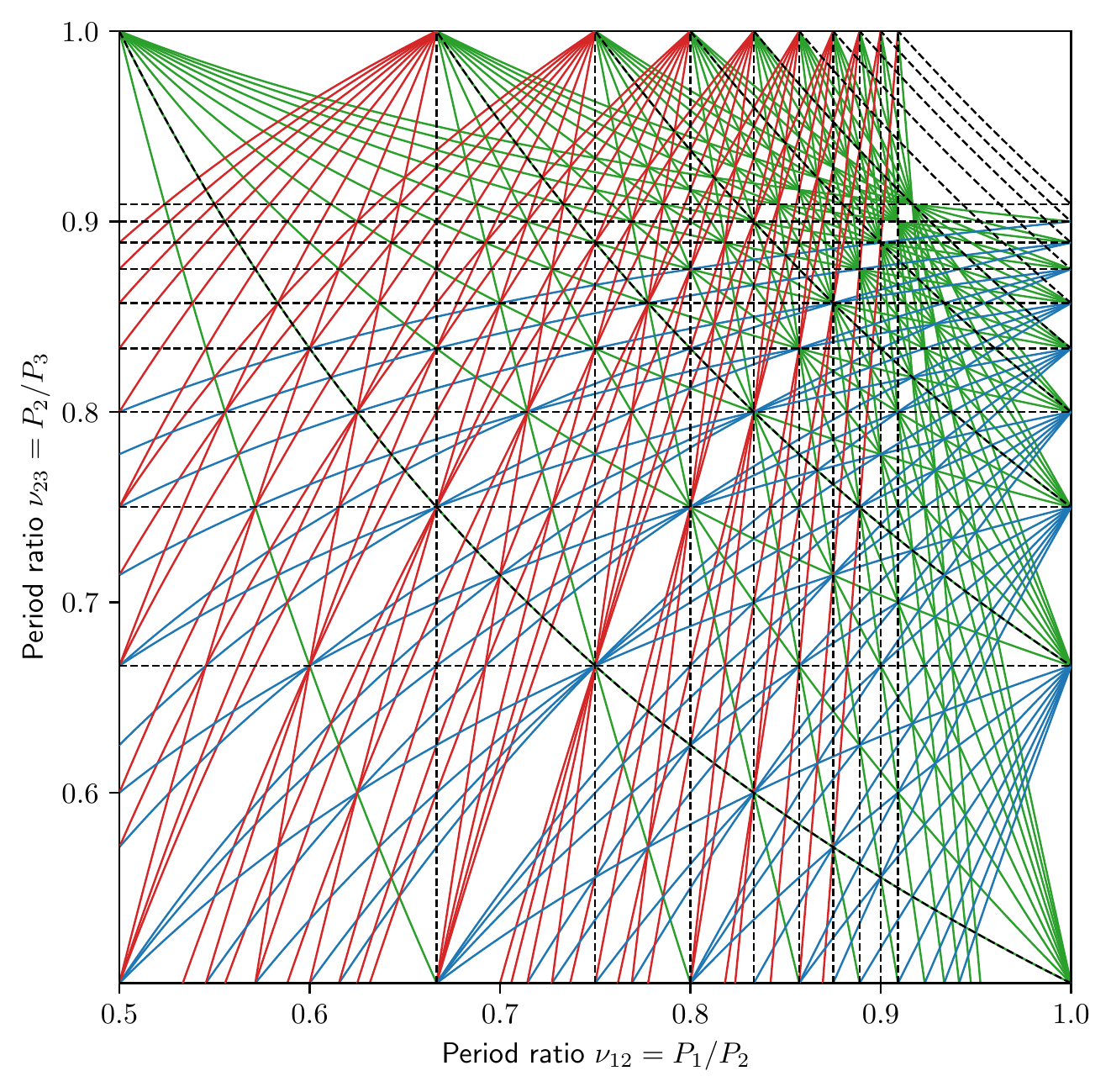}
	\caption{Loci of the first-order three planet resonances with $p+q\leq12$. The colors correspond to the three different families of resonances defined in Eqs. (\ref{eq:blue}, blue), (\ref{eq:red}, red) and (\ref{eq:green}, green). The dashed black lines correspond to first-order two-planet MMRs (obliques lines correspond to resonances between the the first and third planets). One can note that for non tight systems, the majority of the resonances belong to the "red" family (eq. \ref{eq:red}).\label{fig:reslociorder1}}
\end{figure}
Note that the configuration $k_1<0, k_3>0$ is not a solution since the period ratios are smaller than 1.
We plot the loci of the first-order three-planet resonances on Figure \ref{fig:reslociorder1} for $p+q\leq12$.
The colors are associated with the three different families, blue curves correspond to \eqref{eq:blue}, red curves to \eqref{eq:red} and green curves to \eqref{eq:green}.
The blue and red networks are parallel to the zeroth order network while the green network is restricted to the region where the planet pairs are tight. Indeed, all the green resonances with a given $k=p+q$, are outside the square $(0,k/(k+1))\times(0,(k-1)/k))$, which means that for large indexes they become irrelevant.
Nevertheless, one should also note that for $p=q+1$, the corresponding green resonance is actually a two planet MMR between planet 1 and 3.

An important difference with the case of zeroth order resonances treated in \citep{Petit2020a}, is that the resonance locus equation depends specifically on the two integers $k_1$ and $k_3$ rather than on their ratio. 
This property forbids to create a continuous coordinate constant on the resonance loci.
As a result, some simplifications possible in the zeroth-order case are no longer possible.

\begin{figure}
	\centering
	\includegraphics[width=0.85\linewidth]{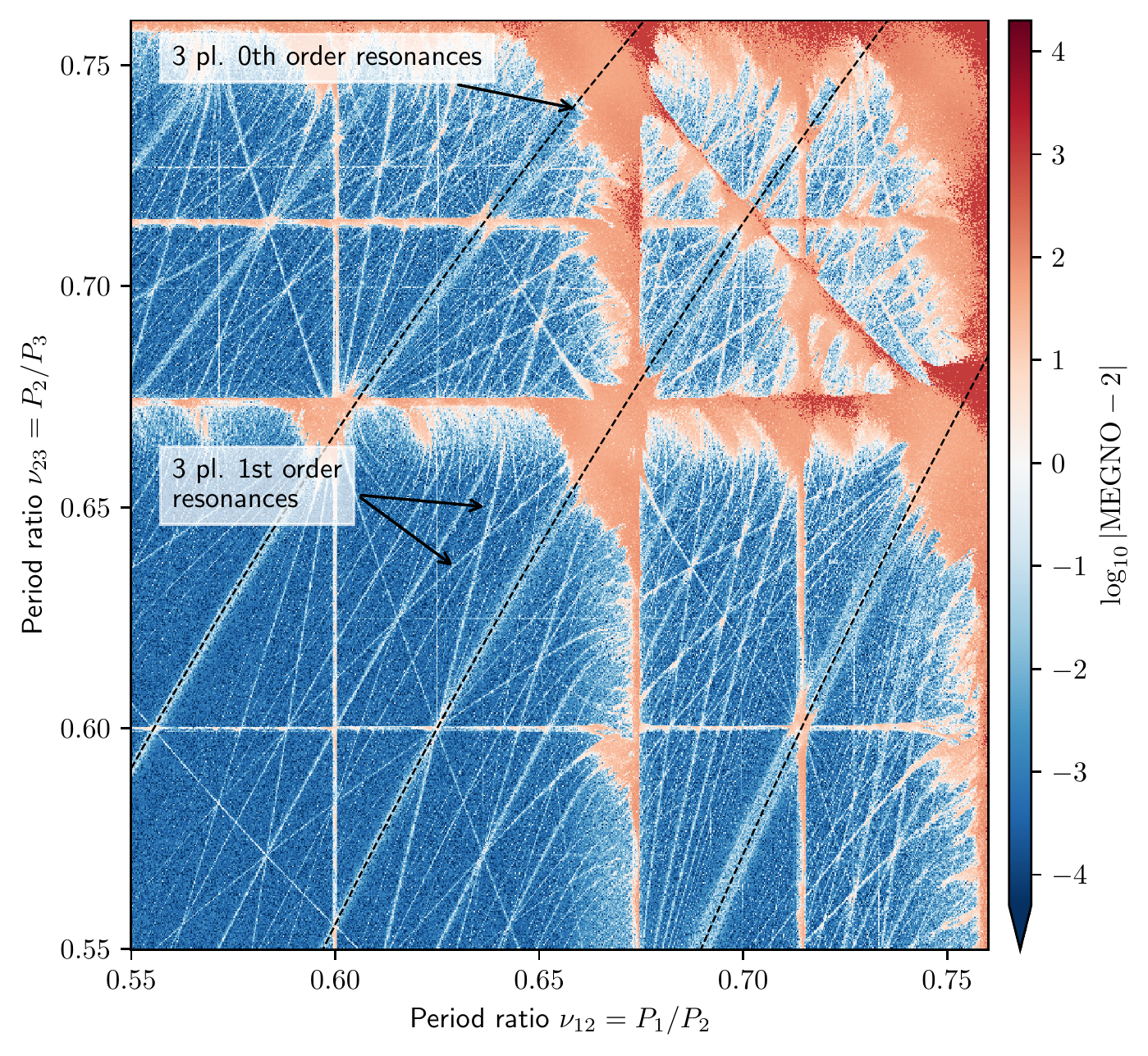}
	\caption{Dynamical map for three planet of masses $10^{-4} M_\odot$ on initially coplanar and circular orbits measured thanks to the MEGNO. Non chaotic regions of the phase space have small values of $|\textrm{MEGNO}-2|$ while it grows to infinity for chaotic systems. First-order three-planet MMRs appears as thin moderately chaotic lines as described by the legend. Zeroth-order three-planet MMRs are larger but harder to spot, hence dashed black lines were added to highlight them.\label{fig:dynmapcirc}}
\end{figure}

Despite their small width, we can observe the resonant network onto dynamical maps. Similarly to \citet{Gallardo2016} or \citet{Charalambous2018}, I perform a series of simulations of three equal masses planet systems.
The initial period ratios $\perrat{12}$ and $\perrat{23}$ span $(0.55,0.76)$ forming a grid of $800\times800$ systems.
Planet masses are $10^{-4} M_\odot$, the orbits are coplanar and initially circular. The innermost planet semi-major axis is 1 au, the star mass $1 M_\odot$ and the starting mean longitudes are random.
The masses and the period ratios correspond to typical spacings for exoplanet multiple systems.
The simulations are integrated for $10^4$ yrs using the integrator \texttt{WHFast} \citep{Rein2015a} from the \texttt{REBOUND} library \citep{Rein2012a}.
I measure the chaotic nature of the dynamics thanks to the Mean Exponential Growth Factor of Nearby Orbits \citep[MEGNO,][]{Cincotta2003}.
The MEGNO oscillates around 2 for quasi-periodic orbits and grow to infinity for chaotic orbits.

The dynamic map is displayed on Figure \ref{fig:dynmapcirc}. One can spot the two-planet MMRs 3:2 and 4:3, responsible for the largest chaotic regions.
The largest zeroth-order three-planet resonances, are highlighted as dashed black lines. Beyond these features, most of the obliques lines showing weak chaos correspond to first-order three-planet resonances. 

\section{Integrable first-order three-planet MMRs Hamiltonian}
\label{sec:model}

\subsection{Second-order averaging}
\label{sec:averaging}

We consider a system of three planets  of masses $m_1,m_2$ , and $m_3$ orbiting a star of mass $m_0$.
The canonical positions $\vec{r}_j$ and momenta $\vec{\tilde{r}}_j$ are expressed in canonical heliocentric coordinates \citep{Poincare1905,Laskar1991}.
The orbits are assumed to be close to circular and coplanar. 
Let the semi-major axes $a_j$, the eccentricities $e_j$, the mean longitudes $\lambda_j$  , and the periapses longitude $\varpi_j$ be the orbital elements defining the orbits.
A set of canonical coordinates for the system is given by the modified Delaunay coordinates \citep[\eg][]{Laskar1991}:
\begin{align}
\Lambda_j &= m_j\sqrt{\mu a_j},  &\lambda_j,\nnb
\AMD_j & = \Lambda_j\left(1-\sqrt{1-e_j^2}\right), &-\varpi_j,\label{eq:Delaunay}
\end{align}
where $\mu = \Gr m_0$ and $\Gr$ is the gravitational constant.
We note that the gravitational parameter $\mu$ is the same for all three planets as in \cite{Laskar2017}.
This is possible by considering the so-called democratic-heliocentric formulation of the planetary Hamiltonian \citep[\eg][]{Morbidelli2002}.
The couples of variables $(\AMD_j,-\varpi_j)$ can also be replaced by their associated complex variables\footnote{$x_j$ are the canonical momenta and $-\i\bar{x}_j$ the conjugated positions}
\begin{align}
x_j &= \sqrt{\AMD_j}\EXP^{\i \varpi_j},& -\i\bar{x}_j.
\label{eq:Poincare}
\end{align}
For small eccentricities, we have ${x_j\simeq \sqrt{\Lambda_j/2}e_j\EXP^{\i \varpi_j}}$. The system total angular momentum $\AM$ and Angular Momentum Deficit \citep[AMD][]{Laskar1997} $C$ are given by
\begin{equation}
\AM = \sum_{j=1}^{3}(\Lambda_j-\AMD_j) \quad \text{and}\quad \AMD= \sum_{j=1}^{3}\AMD_j.
\label{eq:AM-AMD}
\end{equation}

The Hamiltonian $\H$ describing the dynamics can be split into an integrable part  
\begin{equation}
\H_0 = \sum_{j=1}^{3}\frac{\|\vec{\tilde{r}}_j\|^2}{2m_j}-\frac{\mu m_j}{r_j} = -\sum_{j=1}^3 \frac{\mu^2m_j^3}{2\Lambda_j}
\label{eq:HKeplerian}
,\end{equation} 
describing the motion on unperturbed Keplerian orbits, and a perturbation, \begin{equation}
\epsilon \H_1  = -\sum_{j<j'}\frac{\Gr m_jm_{j'}}{|\vec{r}_j-\vec{r}_{j'}|}  + \frac{1}{2m_0}\left\|\sum_{j=1}^3\vec{\tilde{r}}_j\right\|^2
\label{eq:Hpert},
\end{equation}
describing the planet interactions. Here, $\epsilon$ is a dimensionless parameter of the order of the planet-to star-mass-ratio to reflect the scale difference between the two parts of the Hamiltonian.
I limit my study to term at first-order in eccentricity and inclination. This assumption justifies only treating the planar case since inclination terms only appear at order two in the inclinations. The relevant terms of the perturbing Hamiltonian $\epsilon\H_1$ take the form \citep{Laskar1995,Murray1999}
\begin{align}
\epsilon\H_1 =& \phantom{+}\sum_{1\leq i<j\leq 3}\sum_{l\in\Z} (W_{ij}^l +S_{ij}^l) \EXP^{\i l(\lambda_i-\lambda_j)}\label{eq:exprH1}\nnb
&+\sum_{1\leq i<j\leq 3}\sum_{l\in\Z} \left(V_{ij,<}^{l}x_i+V_{ij,>}^{l}x_j\right) \EXP^{\i (l\lambda_i-(l+1)\lambda_j)} +c.c.\\
&+ \sum_{1\leq i<j\leq 3}\sum_{l\in\Z} U_{ij}^l \EXP^{\i (l\lambda_i-(l+2\lambda_j)}  +c.c.\nnb
&+ \sum_{1\leq i,j\leq 3} \frac{m_in_j\Lambda_j}{2m_0\sqrt{\alpha_{ij}}} \EXP^{\i(\lambda_i-\lambda_j)}\left( 1 + \sqrt{\frac{2}{\Lambda_i}}x_i\EXP^{-\i\lambda_j}+\sqrt{\frac{2}{\Lambda_j}}\bar x_j\EXP^{\i\lambda_i}\right) +c.c.,\nonumber
\end{align}
where, in the last three sums, $c.c.$ designates the complex conjugate of the term, and
\begin{align}
W_{ij}^l &=  -\frac{m_in_j\Lambda_j}{2m_0} \lapc{1/2}{l}{\alpha_{ij}},\label{eq:Wijl}\\
V_{ij,<}^{l} &= \frac{m_in_j\Lambda_j}{2m_0} \sqrt{\frac{2}{\Lambda_i}} \left(l+1+\frac{\alpha_{ij}}{2}\dpart{}{\alpha}\right)\lapc{1/2}{l+1}{\alpha_{ij}},\label{eq:Vijl<}\\
V_{ij,>}^{l} &= -\frac{m_in_j\Lambda_j}{2m_0} \sqrt{\frac{2}{\Lambda_j}} \left(l+\frac{1}{2}+\frac{\alpha_{ij}}{2}\dpart{}{\alpha}\right)\lapc{1/2}{l}{\alpha_{ij}},\label{eq:Vijl>}\\
S_{ij}^l &= -\frac{m_in_j\Lambda_j}{m_0}\left(f_2\frac{x_i\bar x_i}{\Lambda_i}+f_2\frac{x_j\bar x_j}{\Lambda_j}+f_{10}\frac{x_i\bar{x}_j+\bar{x_i}x_j}{\sqrt{\Lambda_i\Lambda_j}}\right)\label{eq:Sij}\\
U_{ij}^l &= -\frac{m_in_j\Lambda_j}{m_0}\left(f_{45}\frac{x_i^2}{\Lambda_i}+f_{53}\frac{x_j^2}{\Lambda_j}+f_{49}\frac{x_ix_j}{\sqrt{\Lambda_i\Lambda_j}}\right)\label{eq:Uij}
\end{align}
where $\alpha_{ij}=a_i/a_j$ and $ \lapc{s}{l}{\alpha}$ are the Laplace coefficients \citep[see \eg][]{Laskar1995}. 
$S_{ij}^l$ and $U_{ij}^j$ are the terms at second order in eccentricities and the coefficients $f_k$ are taken from \cite{Murray1999} and defined explicitly in Appendix \ref{app:2ndorder}.
I emphasize that while the dependency in $x_j$ is written explictly for first-order terms, I choose to keep it implicit in the second order terms as this helps keep the computations clearer.
The inclusion of the second order terms is necessary since they create linear terms in eccentricity during the averaging process described below in Section \ref{sec:coefficients}.
The last sum in \eqref{eq:exprH1} corresponds to the indirect terms due to the star's reflex motion. However, as in the two planet case, at first-order in eccentricity, the indirect terms are only relevant for resonances where $p= \pm 1$ or $q= \pm 1$.
As a result, I will not consider the indirect part in the computation of the resonance model\footnote{I still consider them when averaging over the non resonant angles.}.

In the unperturbed case, the system is said to be in a three-body MMR if the mean motions 
\begin{equation}
n_j =\dot{\lambda}_j =  \dpart{\H_0}{\Lambda_j} = \frac{\mu^2m_j^3}{\Lambda_j^3}
\label{eq:meanmotion}
\end{equation}
verify an equation of the form $k_1n_1+k_2n_2+k_3n_3 = 0$.
The sum $k=k_1+k_2+k_3$ is the {`order'} of the resonance.
The sum $K = |k_1| + |k_2| + |k_3|$ is the {`index'} of the resonance.
Because of the d'Alembert rules \citep[\emph{e.g.}][]{Morbidelli2002}, the leading order term in the perturbation is of order $k$ in eccentricity.

Three-planet resonances emerge in the perturbative Hamiltonian after canonical transformations that aim at removing the dependency at first-order in $\epsilon$ of $\epsilon\H_1$ into the fast, non-resonant angles $\lambda_j$.
Such averaging is possible if the system is situated 'far enough' from any two-planet MMR using the classical approach from perturbation theory, the Lie series method \citep{Deprit1969}.
In order to perform the averaging and compute the resonant coefficients, I follow the same method and notations as in \citep{Petit2020a} and I refer the reader to this work for a more extensive description.
Nevertheless is it convenient to recall the notations used.
I note $\epsilon\chi_1$ the generating Hamiltonian, solution of the homological equation
\begin{equation}
	\left\{\epsilon \chi_1,\H_0\right\} +\epsilon \H_1 = \epsilon\bar{\H}_1,
	\label{eq:homological}
\end{equation}
where $\epsilon\bar{\H}_1$ is the average of $\epsilon\H_1$ over the mean longitudes $\lambda$ and $\{\cdot,\cdot\}$ is the Poisson bracket\footnote{I use the convention $\{f,g\} = \sum_j \left(\dpart{f}{p_j}\dpart{g}{q_j}- \dpart{f}{q_j}\dpart{g}{p_j}\right)$ where $(\vec{p},\vec{q})$ is a set of conjugated coordinates.}.
Noting $\hk=\hk(\vec{\Lambda},\mathbf x,\bar{\mathbf x}),$ the complex Fourier coefficients of $\H_1$ with respect to the mean longitudes,
$\epsilon\chi_1$ has for expression
\begin{equation}
\epsilon\chi_1 = \epsilon\sum_{\vec{k}\neq0} \frac{\hk}{\i\vec{k}\cdot\vec{n}}\EXP^{\i\vec{k}\cdot\boldsymbol\lambda}.
\label{eq:chi}
\end{equation}
Due to the expression of $\epsilon\H_1$ given in eq. \eqref{eq:exprH1}, the denominators $\vec{k}\cdot\vec{n}$ are of the form $k_jn_j+k_{j'}n_{j'}$ (in particular they only include terms relative to single pairs of planets) and are not `too small' because I assume the system to be far from two-planet MMRs. Thus the formal series \eqref{eq:chi} is well defined; one can stop the summation at indices $\vec{k}$ of sufficiently high order so that the remaining Fourier terms in $\epsilon\H_1$ have sizes smaller than $\epsilon^2$, which is ensured by the exponential decay of the Fourier coefficients. 

The averaged coodinates (noted with a superscript 1) are obtained by applying the transformation defined by the flow at time -1 of the Hamiltonian function $\epsilon\chi_1$ to the osculating coordinates
\begin{equation}
	(\Lambda^1_j,C^1_j,\lambda^1_j,-\varpi^1_j) = \exp\left(-\left\{\epsilon\chi_1,\cdot\right\}\right) (\Lambda_j,C_j,\lambda_j,-\varpi_j).
\end{equation}
The new Hamiltonian is 
\begin{equation}
	\H^1 = \H_0+ \epsilon\bar \H_1 + \epsilon^2 \H_2 +O(\epsilon^3),
\end{equation}
where the second-order term in $\epsilon$ can be expressed as 
\begin{equation}
\epsilon^2 \H_2 =\frac{1}{2}\left\{\epsilon \chi_1, \epsilon \H_1 + \epsilon \bar \H_1\right\}.
\label{eq:H2full}
\end{equation}
$\epsilon^2 \H_2$ still contains terms depending on fast angles.
The study of a particular three-planet MMR can be done by a second averaging over the non-resonant angles.
In theory, this results in another change of coordinates, which are $\epsilon^2$ close to the first-order averaged coordinates.
In practice I drop these $\epsilon^2$-order corrections to the averaged variables.
I also drop the terms of order $\epsilon^3$ and greater.
To keep the notations light, from now on, unless specified differently I will only consider the averaged coordinates and drop the superscripts. 

\subsection{Change of variables to resonant angles}

The three families of resonances Eqs. (\ref{eq:blue}--\ref{eq:green}) can be treated within the same framework.
As in the two-planet case, there are three resonant angles (one per planet) associated with a first-order three-planet resonance.
Indeed, due to d'Alembert rules, the terms in the Hamiltonian must depend on an angle of the form $\varphi_j = k_1\lambda_1+k_2\lambda_2+k_3\lambda_3-\varpi_j$, where $j=1,2,3$.
Following the same strategy as in \citep{Delisle2012a,Petit2017}, I make a linear, canonical change of coordinates that keeps the highest degree of symmetry. The new set of angles is
\begin{subequations}
\begin{align}
\theta_1 &= k_1(\lambda_2-\lambda_1),\\
\theta_3 &= k_3(\lambda_2-\lambda_3),\\
\theta_\AM &= k_1\lambda_1+k_2\lambda_2+k_3\lambda_3,\\
\varphi_j &= \theta_\AM-\varpi_j\label{eq:resangle}
\end{align}
\end{subequations}
which give for actions
\begin{subequations}
\begin{eqnarray}
	\scalefactor_1 &= \frac{k_1-1}{k_1}\Lambda_1+\Lambda_2+\Lambda_3\label{eq:scale1}\\
	\scalefactor_3 &= \Lambda_1+\Lambda_2+\frac{k_3-1}{k_3}\Lambda_3\label{eq:scale3}\\
	\AM &= \Lambda_1+\Lambda_2+\Lambda_3- \AMD\label{eq:AM}\\
	\AMD_j,&
\end{eqnarray}
\end{subequations}
where $\AMD$ is the total AMD. $\AM$ is the system total angular momentum and by analogy with the two-planet case \citep{Michtchenko2008}, $\scalefactor_1$ and $\scalefactor_3$ are two spacing parameters.
I also define the set of complex coordinates $(-\i\bar{x}'_j,x_j')$ associated with $(C_j,\phi_j)$ where
\begin{equation}
	x'_j = \sqrt{C_j}\EXP^{\i \phi_j}.
\end{equation}

The angles $\theta_1$, $\theta_3$ and $\theta_\AM$ are fast angles for the resonance motion. The second-order formal averaging over the fast angles removes them form the Hamiltonian that thus only depend the angles $\varphi_j$.
As a result, $\scalefactor_1,\scalefactor_3$ and $\AM$ are constant of the motion for a particular resonance defined by $k_1,k_2$ and $k_3$.
Inverting the action coordinates transformation, we have
\begin{align}
	\Lambda_1 &= k_1(\AM+\AMD-\scalefactor_1),\nnb
	\Lambda_2 &= k_2(\AM+\AMD)+k_1\scalefactor_1+k_3\scalefactor_3,\\
	\Lambda_3 &= k_3(\AM+\AMD-\scalefactor_3).\nonumber
\end{align}
One can note that the variations of the semi-major axes only depend on the variation of the total AMD and not on the individual planet AMDs.
This property is similar to the two-planet case \citep{Delisle2014a,Petit2017}.
In order to make progress, one needs to develop the actions close to the resonant manifold.
In the standard derivation of the integrable model for two planet in MMR \citep[\eg][]{Deck2013,Delisle2014a}, the action variables are renormalized by the scaling parameter. This reflects the fact that the dynamics can be rescaled by the orbital timescale of the system.
One can then write the actions $\Lambda_j=\Lambda_{j,0}+\smallvar{\Lambda_j}$ such that the constant values $(\Lambda_{1,0},\Lambda_{2,0},\Lambda_{3,0})$ verifies the Keplerian resonance relationship.
In our case, there are two scaling factors describing the spacing of the two pairs of planets.
As a result, I skip the renormalization phase but keep the idea to develop around the Keplerian resonance.
Indeed, for a given value of the constants of motion $\scalefactor_{1,0}$ and $\scalefactor_{3,0}$, there is a unique angular momentum $\AM_0=\Lambda_{1,0}+\Lambda_{2,0}+\Lambda_{3,0}$ such that the circular orbit configuration with such semi-major axes verify the resonance relationship\footnote{One verifies this by remarking that the determinant of the Jacobian $|J| = 3\sum_{j=1}^3 \frac{k_j^2n_j^2}{\Lambda_j}$ of the system of equations \eqref{eq:res},\eqref{eq:scale1} and \eqref{eq:scale3} is strictly positive, hence the existence of a unique solution.}.
Noting $\DG = G_0-G$, we can write
\begin{align}
	\Lambda_1 &= \Lambda_{1,0}+k_1(\AMD-\DG),\nnb
	\Lambda_2 &= \Lambda_{2,0}+k_2(\AMD-\DG),\label{eq:Lambdavar}\\
	\Lambda_3 &= \Lambda_{3,0}+k_3(\AMD-\DG).\nonumber
\end{align}

\subsection{Computation of the perturbative coefficients}
\label{sec:coefficients}

We need to select from Eq.~\eqref{eq:H2full}, the terms that remain after the averaging of the non resonant angles. 
Since $\epsilon\bar\H_1$ is independent of $\lambda_j$, the term $\{\epsilon\chi,\epsilon\bar\H_1\}$ only contains terms depending on two planet mean longitudes and does not contribute to the resonant dynamics.
Due to the form of the terms of $\epsilon\H_1$, $\epsilon\bar\H_1$ and $\epsilon\chi_1$, there are twelve terms (and their complex conjugates) contributing to the resonant Hamiltonian of a first-order three-planet MMR. 
For the resonance defined by $k_1$ and $k_3$, four terms are created by the combination of a zeroth order term and a first order term
\begingroup
\allowdisplaybreaks
\begin{subequations}
	\begin{align}
		\left\{ \frac{ -\i W^{k_1}_{12} }{ k_1(n_1-n_2)} \EXP^{\i k_1(\lambda_1-\lambda_2)},\left( V^{k_3-1}_{23,<} \bar{x}_2 + V^{k_3-1}_{23,>} \bar{x}_3 \right)\EXP^{\i (-(k_3-1)\lambda_2+k_3\lambda_3)} \right\}, \\
		\left\{ \frac{ \i\left( V^{k_3-1}_{23,<} \bar{x}_2 + V^{k_3-1}_{23,>} \bar{x}_3 \right) }{ (k_3-1)n_2-k_3n_3} \EXP^{\i (-(k_3-1)\lambda_2+k_3\lambda_3)} ,W^{k_1}_{12}\EXP^{\i k_1(\lambda_1-\lambda_2)}\right\}, \\
		\left\{ \frac{ \i W^{-k_3}_{23} }{ k_3(n_2-n_3)} \EXP^{-\i k_3(\lambda_2-\lambda_3)},\left( V^{-k_1}_{12,<} \bar{x}_1 + V^{-k_1}_{12,>} \bar{x}_2\right)\EXP^{\i (k_1\lambda_1-(k_1-1)\lambda_2)} \right\}, \\
		\left\{ \frac{-\i \left(V^{-k_1}_{12,<} \bar{x}_1 + V^{-k_1}_{12,>} \bar{x}_2\right) }{k_1n_1-(k_1-1)n_2} \EXP^{\i (k_1\lambda_1-(k_1-1)\lambda_2)} ,W^{-k_3}_{23}\EXP^{-\i k_3(\lambda_2-\lambda_3)}\right\}.
	\end{align}
	\label{eq:termsPoissonbracket}
\end{subequations}
\endgroup
For these terms, the Poisson bracket reduces to \(\dpart{}{\Lambda_2}\dpart{}{\lambda_2}-\dpart{}{\lambda_2}\dpart{}{\Lambda_2}\).
Furthermore, a combination of a second order term with a first order term can create a term depending on the resonant angle.
There are eight different terms to consider
\begingroup
\allowdisplaybreaks
\begin{subequations}
	\begin{align}
		\left\{ \frac{ -\i V^{-k_1}_{12,>}\bar x_2 \EXP^{\i (k_1\lambda_1-(k_1-1)\lambda_2)}}{ k_1n_1-(k_1-1)n_2)} ,S_{23}^{-k_3}\EXP^{-\i k_3(\lambda_2-\lambda_3)} \right\}, \\
		\left\{\frac{\i S_{23}^{-k_3}\EXP^{-\i k_3(\lambda_2-\lambda_3)}}{k_3(n_2-n_3)}, V^{-k_1}_{12,>}\bar x_2 \EXP^{\i (k_1\lambda_1-(k_1-1)\lambda_2)} \right\},  \\
		\left\{ \frac{ -\i V^{k_1}_{12,>} x_2 \EXP^{\i (k_1\lambda_1-(k_1+1)\lambda_2)}}{ k_1n_1-(k_1+1)n_2)} ,\bar{U}_{23}^{k_3-2}\EXP^{\i (-(k_3-2)\lambda_2+k_3\lambda_3)} \right\}, \\
		\left\{\frac{\i \bar{U}_{23}^{k_3-2}\EXP^{\i (-(k_3-2)\lambda_2+k_3\lambda_3)}}{(k_3-2)n_2-k_3n_3)}, V^{k_1}_{12,>}x_2 \EXP^{\i (k_1\lambda_1-(k_1+1)\lambda_2)} \right\} , \\
		\left\{ \frac{ \i V^{k_3-1}_{23,<}\bar x_2 \EXP^{\i (-(k_3-1)\lambda_2+k_3\lambda_3)}}{ k_1n_1-(k_1-1)n_2)} ,S_{12}^{k_1}\EXP^{\i k_1(\lambda_1-\lambda_2)} \right\} ,\\
		\left\{\frac{-\i S_{12}^{k_1}\EXP^{\i k_1(\lambda_1-\lambda_2)}}{k_1(n_1-n_2)}, V^{k_3-1}_{23,<}\bar x_2 \EXP^{\i (-(k_3-1)\lambda_2+k_3\lambda_3)} \right\} , \\
		\left\{ \frac{ \i V^{-(k_3+1)}_{23,<} \bar x_2 \EXP^{\i (-(k_3+1)\lambda_2+\lambda_3)}}{(k_3+1)n_2-k_3} ,\bar{U}_{12}^{-k_1}\EXP^{\i k_1(\lambda_1-\lambda_2)} \right\},\\
		\left\{\frac{-\i \bar{U}_{12}^{-k_1}\EXP^{\i k_1(\lambda_1-\lambda_2)}}{k_1(n_1-n_2)}, V^{-(k_3+1)}_{23,<} \bar x_2 \EXP^{\i (-(k_3+1)\lambda_2+\lambda_3)} \right\}. 
	\end{align}
	\label{eq:termsPoissonbracket2}
\end{subequations}
\endgroup
Since I limit myself to an expansion at first-order in eccentricity, the Poisson bracket reduces to either \(\i\dpart{}{x_2}\dpart{}{\bar x_2}\) or \(-\i\dpart{}{\bar x_2}\dpart{}{x_2}\). The terms in Eqs.~\eqref{eq:termsPoissonbracket2} are absolutely necessary as they have a comparable size to the terms defined in Eqs.~\eqref{eq:termsPoissonbracket}.

As in the case of zeroth-order three-planet MMRs, ones need to compute the terms of Eq. \eqref{eq:H2full} that depend only on the angles $\varphi_j$. The expression of \(\epsilon^2\H_{2,\mathrm{res}}\) as a function of \(V_{ij,\lessgtr}^k\), \(W_{ij}^k\), \(S_{ij}^k\) and \(U_{ij}^k\) is given in Appendix \ref{app:gfun}.
After regrouping the terms, the perturbation part of the Hamiltonian can be expressed as
\begin{equation}
	\epsilon^2\H_{2,\mathrm{res}} = \epsilon^2\left(-R_{1<}\sqrt{\frac{2}{\Lambda_1}}x'_1 + (R_{1>}-R_{3<})\sqrt{\frac{2}{\Lambda_2}}x'_2 + R_{3>}\sqrt{\frac{2}{\Lambda_3}}x'_3\right) + c.c.,
\end{equation}
where
\begin{align}
	\epsilon^2 R_{1\lessgtr} & = \frac{m_1m_3n_2\Lambda_2\alpha_{23}}{2m_0^2}g^{k_1,k_3}_{1\lessgtr}(\alpha_{12},\alpha_{23}),\\
	\epsilon^2 R_{3\lessgtr} & = \frac{m_1m_3n_2\Lambda_2\alpha_{23}}{2m_0^2}g^{k_1,k_3}_{3\lessgtr}(\alpha_{12},\alpha_{23}),
\end{align}
and $\lessgtr$ is a shorthand notation for the two symbols $<$ and $>$. $g^{k_1,k_3}_{j\lessgtr}$ are functions  given in Appendix \ref{app:gfun} that are linear combinations of Laplace coefficients and their derivatives and depend only on the semi-major axis ratio. Like the Laplace coefficients, these functions show an exponential decay with $|k_1|$ and $|k_3|$ as well as with the orbital spacing.

\subsection{Generalized Sessin-Henrard transformation}

In the two-planet case, a final canonical transformation is needed to obtain a one degree of freedom, integrable Hamiltonian.
This transformation, introduced by \citet{Sessin1984} and \citet{Henrard1986}, can be interpreted as a two-dimensional rotation of the canonical complex variables $x'_j$.
Because the perturbation is linear in the $x'_j$ variables, I can generalize the Sessin-Henrard transformation by introducing a three-dimensional rotation, creating two additional integrals of motion.
I define the canonical change of variables\footnote{The vector $(r_1,r_2,r_3)$ can be completed into an orthogonal base by any orthogonal base of the perpendicular plane. As a result the particular choice of $y_2$ and $y_3$ is arbitrary.}
\begin{equation}
	\left({\begin{array}{c}
		y_1 \\
		y_2 \\
		y_3
	   \end{array} }\right) = \left({\begin{array}{ccc}
		r_1 & r_2 & r_3 \\
		\frac{r_3}{\sqrt{1-r_2^2}} & 0 & \frac{-r_1}{\sqrt{1-r_2^2}} \\
		\frac{r_2r_1}{\sqrt{1-r_2^2}} & \frac{r_2^2-1}{\sqrt{1-r_2^2}}  & \frac{r_2r_3}{\sqrt{1-r_2^2}} 
	   \end{array} }\right)\left({\begin{array}{c}
		x'_1 \\
		x'_2 \\
		x'_3
	   \end{array} }\right), \label{eq:Sessin-transform}
\end{equation}
where 
\begin{equation}
	r_1 = -\frac{R_{1<}}{R}\sqrt{\frac{2}{\Lambda_1}}, \quad r_2 = \frac{R_{1>}-R_{3<}}{R}\sqrt{\frac{2}{\Lambda_2}}, \quad r_3 = \frac{R_{3>}}{R}\sqrt{\frac{2}{\Lambda_3}},
\end{equation}
and
\begin{equation}
	\epsilon^2R = \frac{m_1m_3n_2\Lambda_2\alpha_{23}}{2m_0^2}\sqrt{\frac{2(g^{k_1,k_3}_{1<})^2}{\Lambda_1}+\frac{2}{\Lambda_2}\left(g^{k_1,k_3}_{1>}-g^{k_1,k_3}_{3<}\right)^2 +\frac{2(g^{k_1,k_3}_{3>})^2}{\Lambda_3}}.
\end{equation}

This transformation is canonical since the rotation matrix is orthogonal. 
I define the associated action-angle coordinates to the complex coordinates $y_j$ as $(I_j,\psi_j)$, such that $y_j = \sqrt{I_j}\EXP^{\i \psi_j}$.
The AMD still verifies $C = y_1\bar{y}_1+y_2\bar{y}_2+y_3\bar{y}_3 = I_1+ I_2+ I_3$. As in the two-planet case, at the first order in eccentricities, the resonant variable is proportional to linear combination of the complex eccentricities with coefficients depending only on the semi-major axis ratios 
\begin{align}
y_1  &\propto \left(-g^{k_1,k_3}_{1<} e_1\EXP^{-\i\varpi_1} +
\left(g^{k_1,k_3}_{1>}-g^{k_1,k_3}_{3<}\right) e_2\EXP^{-\i\varpi_2} 
+ g^{k_1,k_3}_{3>}e_3\EXP^{-\i\varpi_3}\right)\EXP^{\i \theta_\AM}
\end{align}

Finally, I expand at the second order in $(\AMD-\Delta\AM)$ the Keplerian part around the resonance for unperturbed orbits and drop the first order secular terms. The Hamiltonian becomes
\begin{equation}
	\H = -\frac{\K_2}{2} (I_1+I_2+I_3-\Delta\AM)^2 + 2\epsilon^2R\sqrt{I_1}\cos(\psi_1),
	\label{eq:Hamintegrable}
\end{equation}
where
\begin{equation}
	-\frac{\K_2}{2} = -\frac{1}{2}\sum_{j=1}^3 \dpart{^2\H_0}{\Lambda_j^2} k_j^2=-\frac{3}{2}\sum_{j=1}^3 \frac{n_jk_j^2}{\Lambda_j}. %
\end{equation}
Since $\H$ does not depend on $\psi_2$ and $\psi_3$, the actions coordinates $I_2$ and $I_3$ are constant of the motion, up to perturbations of order $\epsilon^3$.

\section{Structure and width of the resonances}
\label{sec:width}

One can recognize in Eq. \eqref{eq:Hamintegrable} the second fundamental model of resonance proposed by \citet{Henrard1983}. 
This Hamiltonian, also called an Andoyer Hamiltonian of degree one appears in the description of first order two planet MMR, Cassini states and various problems of celestial mechanics.
The width of the resonance as well as the dynamics of this integrable Hamiltonian are well-known and I refer to \citet{Ferraz-Mello2007} for a detailed description. 
For the determination of the resonance widths and structure, I will adapt the formalism from \citet{Petit2017}.
The main difference with the two-planet first-order MMR is that the resonant term is of second order in planet-to-star mass ratio.
\begin{figure}
	\includegraphics[width=\linewidth]{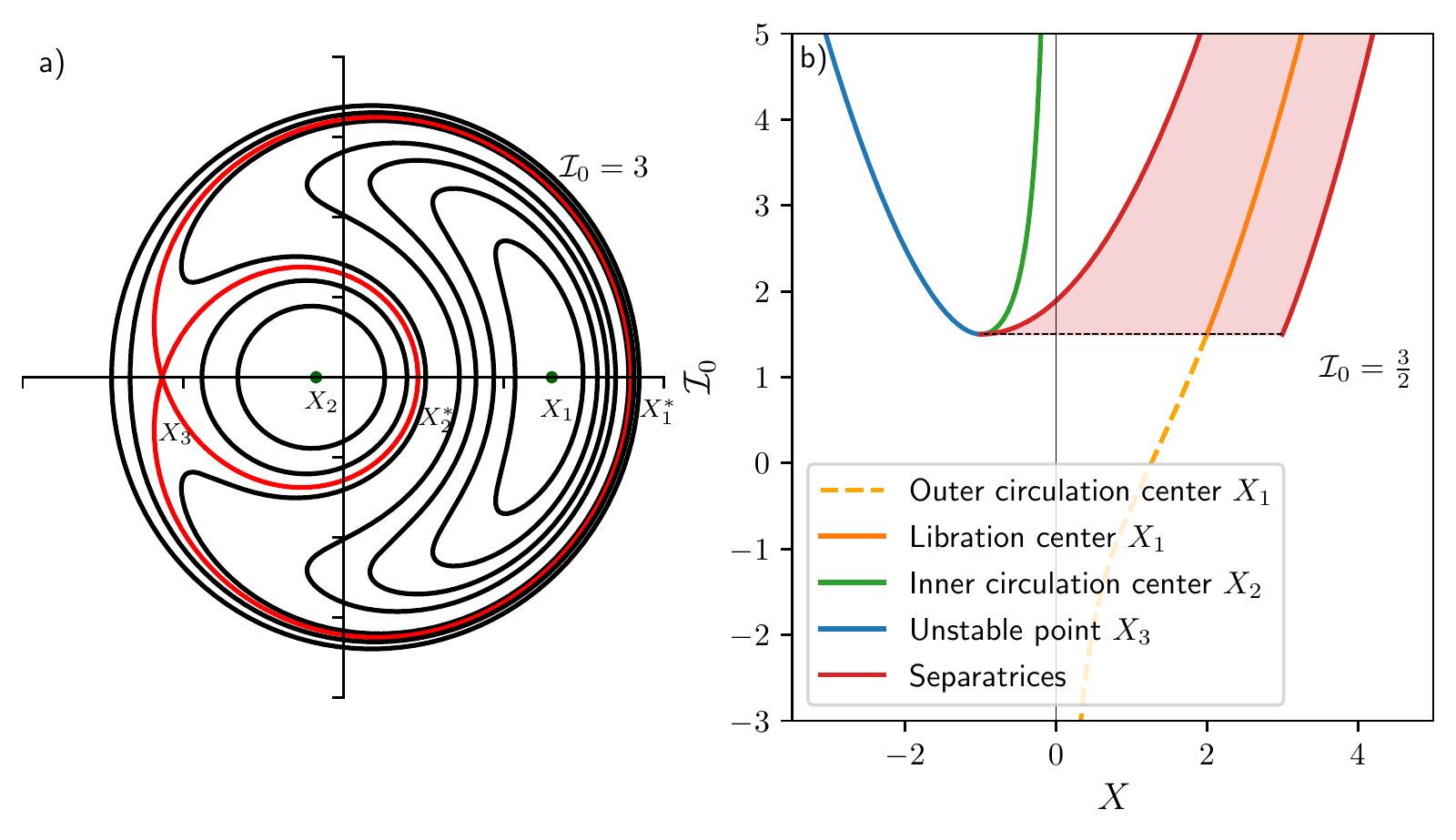}
	\caption{a) Structure of the resonance in the ($X,Y$) plane for $\mathcal{I}_0=3$. b) Position on the $X$-axis of the fixed points and separatrices as a function of $\mathcal{I}_0$. The resonance opens for $\mathcal{I}_0>3/2$ and the resonant configuration are shown in red.}
	\label{fig:resstruct}
\end{figure}
I define 
\begin{equation}
	\epsilon^2\kappa = \frac{\sqrt{2}R}{\K_2},
\end{equation}
as well as $\mathcal{I}_0 = (\epsilon^2\kappa)^{-2/3}(\Delta\AM - I_2-I_3)$, $X = (\epsilon^2\kappa)^{-1/3}\sqrt{2I_1}\cos\psi_1$ and $Y = (\epsilon^2\kappa)^{-1/3}\sqrt{2I_1}\sin \psi_1$.
In these renormalized variables, the Hamiltonian can be written
\begin{equation}
	\H = -\frac{(\epsilon^2\kappa)^{4/3}}{2}\left(\frac{1}{2}(X^2+Y^2)-\mathcal{I}_0\right)^2 + (\epsilon^2\kappa)^{4/3}X.
\end{equation}
The fixed points verify $X^3-2\mathcal{I}_0X-2=0$ and $Y=0$. The cubic equation has three solution if and only if $\mathcal{I}_0>3/2$.
In this case the resonance is open and separatrices exist.
Noting $X_1>X_2\geq X_3$ the abscissas of the fixed points, $X_3$ defines the unstable hyperbolic point where the separatrices meet.
I also note $X_1^*$ and $X_2^*$ the intersections of the separatrices with the $X$-axis.
Since the separatrices are on the same energy level as the unstable fixed point, we have
\begin{equation}
	X_1^* = -X_3+\frac{2}{\sqrt{|X_3|}} \quad\mathrm{and}\quad X_2^*=-X_3-\frac{2}{\sqrt{|X_3|}}.
\end{equation}
In Figure \ref{fig:resstruct}, I plot the structure of the resonance for $\mathcal{I}_0=3$ in the $(X,Y)$ plane as well as the position of the fixed points and separatrix intersections for varying $\mathcal{I}_0$.

The width of the resonance can be expressed implicitly as a function of the minimum value of $I_1$ for a resonant orbit given a value of $\mathcal{I}_0$ \citep{Petit2017}.
Indeed, the resonance width in terms of $I_1$ is
\begin{equation}
	\Delta I_1 = 4(\epsilon^2\kappa)^{2/3}\sqrt{|X_3|}.
\end{equation}
The value of $X_3$ is linked the minimum value of $I_1$ to enter the resonance island thanks to the relationship
\begin{equation}
	C_{\min} = I_1(X_2^*) = \frac{(\epsilon^2\kappa)^{2/3}}{2}\left(-X_3-\frac{2}{\sqrt{|X_3|}}\right)^2.
\end{equation}
There are two limit cases of interest, the case of circular orbits and the large eccentricity regime \citep{Petit2017}.
In the case of circular orbits, the resonance width becomes $(\Delta I_1)_\mathrm{cir} = 4\times2^{1/6}(\epsilon^2\kappa)^{2/3}$.
For eccentric orbits, we have ${X_3\simeq (\epsilon^2\kappa)^{-1/3}\sqrt{2C_{\min}}}$, which leads to $(\Delta I_1)_\mathrm{ecc} = 4\epsilon\sqrt{\kappa \sqrt{2C_{\min}}}$.
One can then obtain the width of the resonance in term of period ratio by using Eqs \eqref{eq:perrat} and \eqref{eq:Lambdavar}
\begin{equation}
	\Delta\perrat{12} = 3\perrat{12}\left|\frac{k_1}{\Lambda_1}-\frac{k_2}{\Lambda_2}\right|\Delta I_1, \quad \mathrm{and} \quad \Delta\perrat{23} = 3\perrat{23}\left|\frac{k_2}{\Lambda_3}-\frac{k_3}{\Lambda_3}\right|\Delta I_1.
\end{equation}

\begin{figure}
	\includegraphics[width=\linewidth]{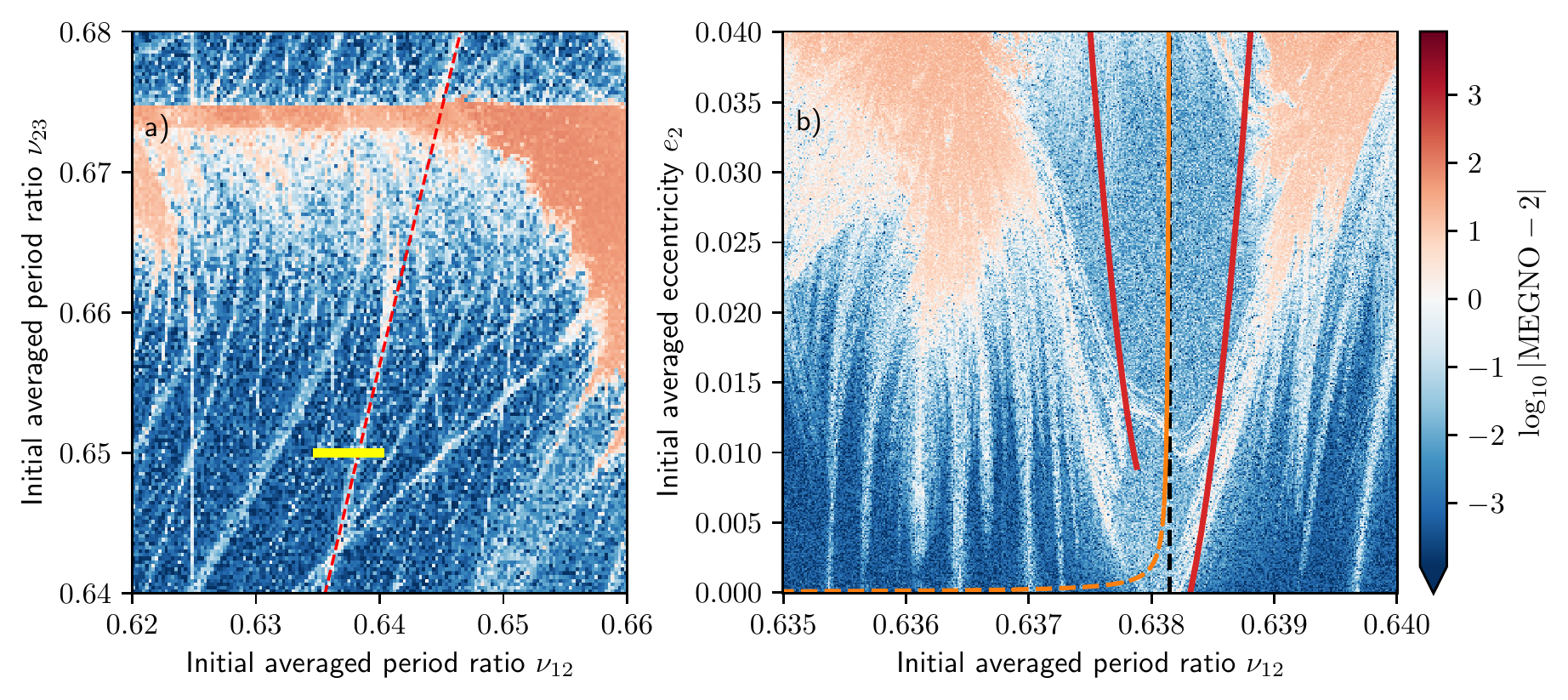}
	\caption{a) Zoom of Figure \ref{fig:dynmapcirc} around the resonance studied in b). The dashed red curve corresponds to the locus of the resonance $-3n_1 + 6 n_2 - 2n_3$ for Keplerian motion and the yellow segment to the extent of the zoomed slice. b) Dynamical map in the plane $(\perrat{12},e_2)$. The red curves corresponds to the inner and outer separatrices, the orange curve to the center of the resonance.\label{fig:vslice32-1}}
\end{figure}

I also compare this analytical result to a dynamical map at the vicinity of the resonance $-3n_1 + 6 n_2 - 2n_3$. I choose to keep $\perrat{23}$ constant while changing $\perrat{12}$ and the eccentricities of the planets. The one-dimensional cut in the period ratio plane is shown on Fig. \ref{fig:vslice32-1}a, that is a zoom of Fig. \ref{fig:dynmapcirc} around the resonance of interest. I initialize a regular grid of $400\times400$ initial conditions with the averaged variables $0.635\leq \perrat{12} \leq 0.640$ and $0\leq e_2 \leq 0.04$.
The averaged outer planet period ratio is kept constant $\perrat{23,0}=0.65$, the orbits are coplanar, the averaged eccentricities $e_1, e_3$ and the mean longitudes are chosen such that $I_2=I_3=0$ and $\arg(y_1)=0$.
As in Figure \ref{fig:dynmapcirc}, the planet masses are $10^{-4} M_\odot$, the systems are integrated $10^4$ orbits and I measure the MEGNO along the trajectory.
The resonance is small with respect to the rapid variations of the osculating coordinates.
Therefore, these averaged initial conditions are converted into osculating orbital elements before the start of the integration.

The dynamical map is shown on Figure \ref{fig:vslice32-1}b.
Qualitatively, the shape of the resonance is similar to the structure observed for two-planet MMR, confirming that the second fundamental model of resonance describes well first-order three-planet MMR.
One can see that the outer separatrix does not go to zero eccentricity on the dynamical map, as predicted by the analytical model.
Similarly the position of the inner separatrix matches well the dynamical map at low eccentricities.

However, the analytical model slightly underestimates the size of the resonance (by roughly a factor 2 for a mean eccentricity \(e_2 \gtrsim 0.01\)).
A possible explanation could be that higher order terms in eccentricity becomes relevant for non circular orbits and could increase the width of the resonance as discussed by \cite{Hadden2019} for the two-planet MMR case.
At higher eccentricities,the width of the resonance is not proportional to \(\sqrt{e_2}\), as it would be in the case of the second fundamental model of resonance.
Despite this slight discrepancy, the qualitative behaviour predicted by the model is correct.

\section{Role of first-order three-planet resonances on the planet dynamics}
\label{sec:applications}

This analytical model for first-order three-planet resonances gives an opportunity to study how they affect exoplanet dynamics.
We saw that while these resonances are weak with respect to low-order two-planet MMR and zeroth-order three-planets MMRs, they can affect the dynamics locally.
In particular the similarities of the dynamics with the two-planet case suggest that their overlap can lead to destabilization of tightly packed systems and that resonance capture is possible in the case of convergent migration, thanks to the open separatrix at low eccentricity.
In this section I show two examples of applications of my analytical model to exoplanet dynamics problems.
While both problems deserve a detailed study, the full development is left for future works. 

\subsection{Contribution to the instability of tightly packed systems}

The motivation for this study emerged while trying to extend the results on the stability of tightly packed systems from \citet{Petit2020a} to eccentric systems.
Indeed, as the eccentricity increases, the first-order three-planet resonance width increases to the point where they should play a role in driving the instability.
The main concept recently developed to treat the question of complex resonance overlap involving varied resonances sizes is the notion of resonance network density or optical depth \citep{Quillen2011,Hadden2018,Petit2020a}.
It consists in comparing the sum of the width of the resonances to the space they occupy.
Under the approximation that the resonances are well distributed, when the total volume of the resonances becomes larger than the available space, large scale chaos can develop.

As explained in section \ref{sec:network}, the main difference between the zeroth-order and first-order resonance networks is that first-order resonances intersect and there is no transformation of coordinates from the period ratios to a coordinate constant onto the resonance loci.

We can nevertheless estimate the density of the first-order three planet resonance network.
For a given resonance described by $k_1$ and $k_3$, I define
\begin{equation}
	f_{k_1,k_3}(n) = k_1n_1 -(k_1+k_3-1)n_2 +k_3n_3.
\end{equation}
The function $f_{k_1,k_3}$ vanishes on the resonance locus.
By construction, we have $\d f_{k_1,k_3}/\d I_1 = -\K_2/2$, which means that the width of the resonance in terms of $f_{k_1,k_3}$ is 
\begin{equation}
	\Delta f = 2\K_2(\epsilon^2\kappa)^{2/3}\sqrt{X_3}.
\end{equation}
The resonant coefficient $\epsilon^2\kappa$ mostly depends on the sum $k_1+k_3$, it thus makes sense to compute the density of the subnetwork of resonances verifying $k_1+k_3$.
We have $f_{k_1+1,k_3-1}-f_{k_1,k_3}=n_1-n_3$, it results that the subnetwork density is
\begin{equation}
	\rho_{k_1+k_3} = \frac{2\K_2(\epsilon^2\kappa)^{2/3}\sqrt{X_3}}{n_2(\perrat{12}^{-1}-\perrat{{23}})}.
\end{equation}
It is in principle possible to obtain an overlap criterion from this expression by summing over the subnetworks.
However, considering the first order resonances alone  is incorrect and one need to take into account the zeroth-order resonances.
Moreover, as the eccentricity increases, more and more two-planet resonances are activated and become the dominant source of chaos.
Due to the complexity of this study, this problem will be the topic of further works.

\subsection{Capture into resonant chains}

Disk-driven planet migration leads to capture into two-planet first-order MMRs \citep[\emph{e.g.}][]{Terquem2007,Cresswell2008,Izidoro2017}.
Contrarily to the general case (like zeroth-order or second and higher order resonances) where the capture into resonance is essentially probabilistic \citep{Henrard1982}, low-eccentricity convergent migration leads to automatic capture into resonance because no separatrix crossing is needed to enter the resonance \citep{Batygin2015}.
The resonance fixed point can however be overstable or the migration can be too fast to allow for capture, depending on the planet masses and migration rate \citep[\eg][]{Ogihara2013,Deck2015}.

Since the three-planet first-order resonances dynamics are similar to the two-planet case, it is natural to consider whether capture into these resonances is possible.
Besides being much weaker than their two-planet counterpart, the main difference regarding resonance capture is the fact that the Keplerian resonance locus is a curve in the period ratio plane rather than a fixed value on a one-dimensional axis.
As a result, after capture, the orbital separation can still tighten, leading to the crossing of other three-planet resonances or even two-planet ones.
While this problem has not been studied to the same level of details than the capture into two-planet MMRs, \citet{Charalambous2018} have shown on the example of systems with configuration similar to Trappist-1 that three-planet first-order resonances can act as guides during migration (their Figures 7 and 8).

\begin{figure}
	\includegraphics[width=\linewidth]{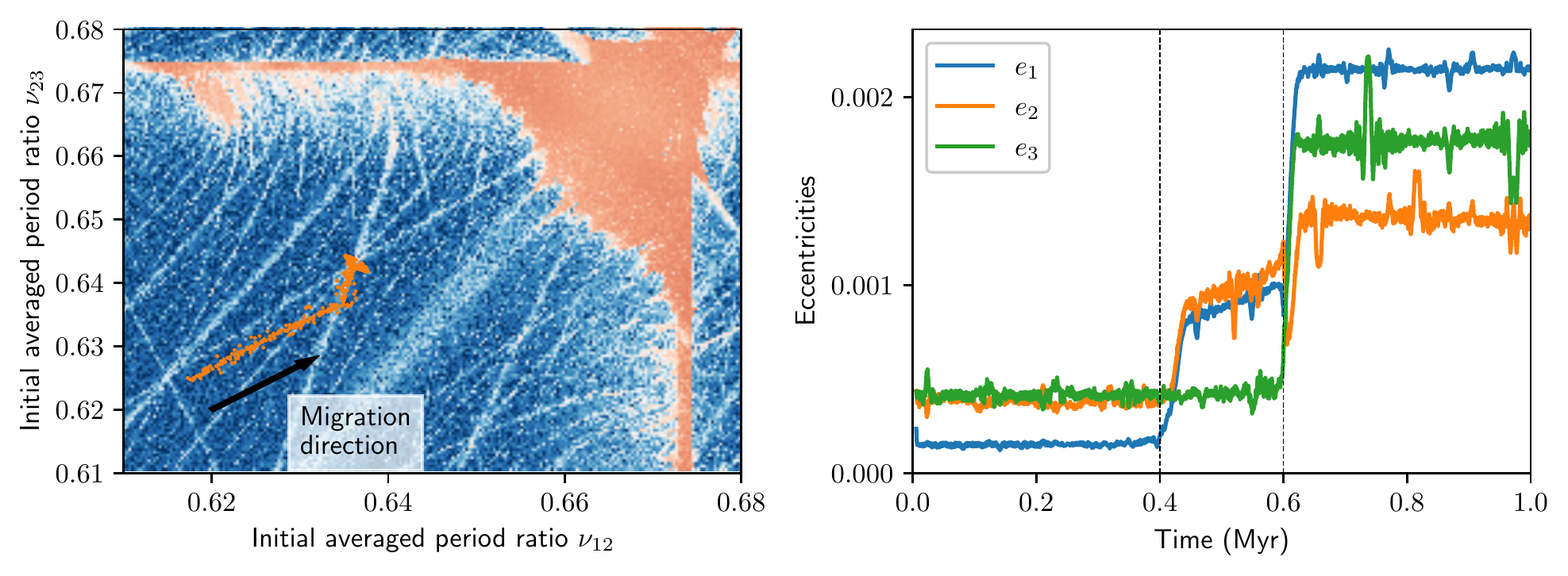}
	\caption{Period ratio and eccentricity evolution due to semi-major and eccentricity damping.
	Left: Evolution in the period ratio plane superposed to a zoom of Figure \ref{fig:dynmapcirc}.
	Right : Eccentricities time series. To improve readability, a moving average over 20 snapshots is applied to remove the rapid fluctuations due to the libration into the resonance.\label{fig:capturegen}}
\end{figure}
\begin{figure}
	\includegraphics[width=\linewidth]{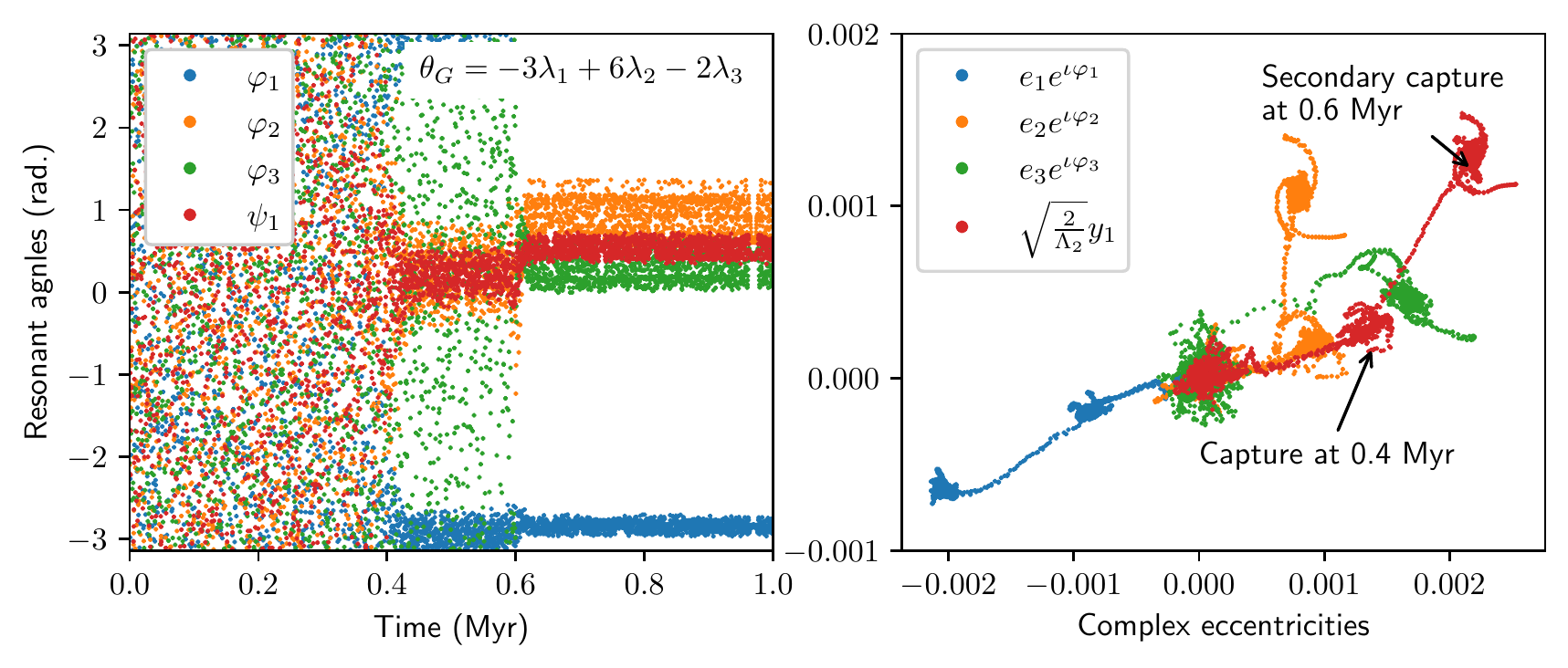}
\caption{Capture into the resonance $-3n_1+6n_2-2n_3$.
	Left: Time series of the three classical resonant angles $\varphi_j$ \eqref{eq:resangle} and of the argument of the resonant variable $\psi_1 = \arg(y_1)$. The capture into the resonance occurs around 0.4 Myr as seen by the libration of $\psi_1$.
	After 0.6 Myr, the system is also trapped into the resonance $2n_2-7n_2+6n_3$ (see Figure \ref{fig:rescap276})
	Right: Evolution of the resonant complex eccentricities as well as the resonant variable $y_1$ \eqref{eq:Sessin-transform} normalized by $\sqrt{2/\Lambda_2}$ in order to be comparable with the eccentricities.
	The moving average is also applied to these variables.\label{fig:rescap362}
	}
\end{figure}

\begin{figure}
	\includegraphics[width=\linewidth]{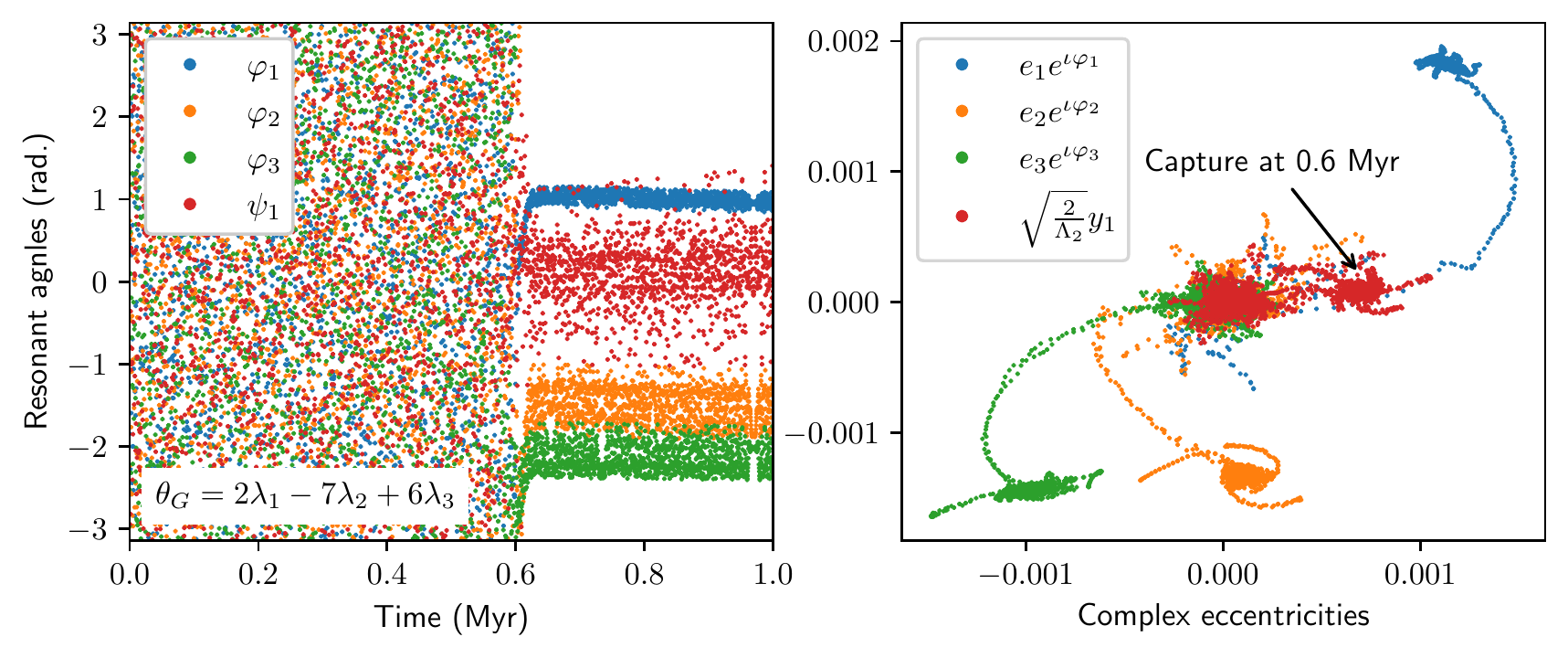}
\caption{
	Capture into the resonance $2n_1-7n_2+6n_3$.
	Left: Time series of the three classical resonant angles $\varphi_j$ \eqref{eq:resangle} and of the argument of the resonant variable $\psi_1 = \arg(y_1)$. The capture into this resonance occurs around 0.6 Myr as seen by the libration of $\psi_1$.
	Right: Evolution of the resonant complex eccentricities as well as the resonant variable $y_1$ \eqref{eq:Sessin-transform} normalized by $\sqrt{2/\Lambda_2}$ in order to be comparable with the eccentricities.
	The moving average is also applied to these variables.\label{fig:rescap276}
	}
\end{figure}

In order to illustrate the capture mechanism, I integrate for 1 Myr a three $10^{-4} M_\odot$ equal-mass planet system with semi-major axis and eccentricity damping implemented thanks to \texttt{REBOUNDx} \citep{Tamayo2019}.
Since the resonances are very weak, it is necessary to adopt very slow damping timescales so that capture is possible.
We follow the constant timescale damping prescription from \citet{Goldreich2014}.
The eccentricity damping timescale is $\tau_e=10^3$ yr for all planets.
The semi-major axis damping timescales are $\tau_{a,1} = +\infty$, $\tau_{a,2} = 20$ Myr and $\tau_{a,3} = 10$ Myr, so that the differential migration rate between the planets is equal to 20 Myr.
3,000 snapshots are recorded and the semi-major axis and complex eccentricities are converted from osculating to averaged variables.

The evolution of the period ratios and eccentricities are plotted in Figure \ref{fig:capturegen}.
Due to the rapid evolution of the eccentricities within the resonances, a moving average over 20 snapshots was applied.
In the period ratio space, we see that the system evolves towards the resonance $-3n_1+6n_2-2n_3$ and reaches it around $t=0.4$ Myr.
Then the system evolves along the resonance locus and stops at 0.6Myr when the resonance $-3n_1+6n_2-2n_3$ crosses the resonance $2n_1-7n_2+6n_3$.
The breaks into the evolution can also be seen onto the eccentricities time series.
We see that $e_1$ and $e_2$ first rise after the first capture before an increase to a constant value after the secondary capture.
We can also see that the resonance angles start to librate after the captures in Figures \ref{fig:rescap362} and \ref{fig:rescap276}.
It should be noted that in Figure \ref{fig:rescap362}, $\psi_1$ librates after the first capture while $\varphi_3$ is not librating.
The capture can also be observed onto the complex eccentricity plane as shown on the right panels of Figures \ref{fig:rescap362} and \ref{fig:rescap276}.
Since the angles $-3\lambda_1+6\lambda_2-2\lambda_3-\varpi_j$ and $2\lambda_1-7\lambda_2+6\lambda_3-\varpi_j$ both librate, the zeroth-order angle $5\lambda_1-13\lambda_2+8\lambda_3$ also librates.
Indeed, the zeroth-order resonance passes through the intersection point of the two first-order resonances.
It is probable that it helps stabilizing the configuration.

This example shows that capture is possible into these resonances.
However, the resonance itself does not stop the migration because it is not a hard barrier in the period ratio space. An intersection of resonances can however lead to capture at fixed period ratios.
Importantly, these results were obtained in a very controlled environment with ad-hoc migrations timescales, similarly to the experiments of \cite{Charalambous2018}.
For a faster migration or smaller planets, the resonance width will be too small to allow for capture as the system will swipe through in less than a libration period.
The necessary conditions leading to such capture may thus not be present in the context of a realistic formation scenario.

\section{Conclusion}

I have shown that the dynamics of all first-order three-planet resonances can be approximated by a novel integrable model using the same strategy than the first-order two planet MMRs.
After a second-order averaging in the planet masses over the non-resonant angles,
the dynamics can be reduced to a second fundamental model of resonance \citep{Henrard1983} thanks to a generalized Sessin-Henrard transformation of the eccentricity variables \citep{Sessin1984,Henrard1986}.
My model is the first analytical solution in the presence of three massive planets instead of two-planet and a test particle.
From there, I derived the width and the shape of the resonances as a function of the period ratios, planet masses and eccentricities of the planet.
In the context of an isolated resonance, the predicted analytical width is within a factor 2 of the observed width on numerical simulations.
The analytical model matches really well the width of the resonance for close to circular orbits.
I showed that the analytical width could be used in future work in order to predict more accurately the lifetime of systems with moderate eccentricities.
The resonance structure of first-order three-planet MMR makes capture into these resonances possible, although it is not clear which regime can most likely lead to it as my results were obtained in a controlled environment.

As shown on Figure \ref{fig:dynmapcirc}, isolated first-order three-planet resonances are mostly important just outside from the two-planet and three-planet MMR overlap limits \citep{Hadden2018,Petit2020a}.
In the context of exoplanets, this typically corresponds to compact systems with a planet to star mass ratio of the order $10^{-4}$ with period ratios between 1.33 and 1.7.
Observationally, it is not possible to confirm directly if a system is trapped into such resonance because the resonant angles depend on the pericentre longitudes that are poorly constrained in general, particularly for close to circular systems.
However, in the context of very tight systems, a dynamical study can restrict the stable configurations to the resonant island, suggesting that such resonances exist.
In particular, such a scenario was proposed in the context of Kepler 11 \citep{Migaszewski2012}.
The possibility that exoplanet systems are trapped into first-order three-planet MMRs deserves more investigation and will be the topic of future works.

 \begin{acknowledgements}
	This work is by supported by the Royal Physiographic Society of Lund through the Fund of the Walter Gyllenberg Foundation (number 40730).
 \end{acknowledgements}

\section*{Conflict of interest}
The author declare that they have no conflict of interest.

\appendix
\section{Terms at second-order in eccentricity}
\label{app:2ndorder}

The second-order terms in eccentricity in the Hamiltonian takes the form
\begin{align}
S_{ij}^l &= -\frac{m_in_j\Lambda_j}{m_0}\left(f_2\frac{x_i\bar x_i}{\Lambda_i}+f_2\frac{x_j\bar x_j}{\Lambda_j}+f_{10}\frac{x_i\bar{x}_j+\bar{x_i}x_j}{\sqrt{\Lambda_i\Lambda_j}}\right)\label{eq:appSij},\\
U_{ij}^l &= -\frac{m_in_j\Lambda_j}{m_0}\left(f_{45}\frac{x_i^2}{\Lambda_i}+f_{53}\frac{x_j^2}{\Lambda_j}+f_{49}\frac{x_ix_j}{\sqrt{\Lambda_i\Lambda_j}}\right)\label{eq:appUij},
\end{align}
where the functions $f_k$ are defined in Appendix B of \cite{Murray1999} and are function of the semi-major axis ratio and the index $l$ only. 
I use a different convention for the arguments nomenclature, for example \cite{Murray1999} defines the arguments as \((l-1)\lambda_1-l\lambda_2+\varpi_1\) whereas I define such a term as \(l\lambda_1-(l+1)\lambda_2+\varpi_1\). 
As a result I adapt the expressions of \(f_k\) to reflect this change of convention.
I use
\begin{subequations}
\begin{align}
f_2^l &= \frac{1}{8}\left(-4l^2+2\alpha\dpart{}{\alpha} +\alpha^2\dpart{^2}{\alpha^2}\right)\lapc{1/2}{l}{\alpha},\\
f_{10}^l &= -\frac{1}{4}\left(-4l^2-6l-2+2\alpha\dpart{}{\alpha} +\alpha^2\dpart{^2}{\alpha^2}\right)\lapc{1/2}{l+1}{\alpha},\\
f_{45}^l &= \frac{1}{8}\left(4l^2+11l+6+(4l+6)\alpha\dpart{}{\alpha} +\alpha^2\dpart{^2}{\alpha^2}\right)\lapc{1/2}{l+2}{\alpha},\\
f_{49}^l &= -\frac{1}{4}\left(4l^2+10l+6+(4l+6)\alpha\dpart{}{\alpha} +\alpha^2\dpart{^2}{\alpha^2}\right)\lapc{1/2}{l+1}{\alpha},\\
f_{53}^l &= \frac{1}{8}\left(4l^2+6l+4+(4l+6)\alpha\dpart{}{\alpha} +\alpha^2\dpart{^2}{\alpha^2}\right)\lapc{1/2}{l}{\alpha}.
\end{align}
\end{subequations}

\section{Derivatives of the perturbation terms}
\label{app:derivatives}

As described in the main text, the zeroth- and first-order factors in eccentricity of the Hamiltonian can be expressed as
\begin{subequations}
\begin{align}
	W_{ij}^l &=  -\frac{m_in_j\Lambda_j}{2m_0} \lapc{1/2}{l}{\alpha_{ij}},\label{eq:appWijl}\\
	V_{ij,<}^{l} &= \frac{m_in_j\Lambda_j}{2m_0} \sqrt{\frac{2}{\Lambda_i}} \left(l+1+\frac{\alpha_{ij}}{2}\dpart{}{\alpha}\right)\lapc{1/2}{l+1}{\alpha_{ij}},\label{eq:appVijl<}\\
	V_{ij,>}^{l} &= -\frac{m_in_j\Lambda_j}{2m_0} \sqrt{\frac{2}{\Lambda_j}} \left(l+\frac{1}{2}+\frac{\alpha_{ij}}{2}\dpart{}{\alpha}\right)\lapc{1/2}{l}{\alpha_{ij}}.\label{eq:appVijl>}
	\end{align}
\end{subequations}
The derivatives of $W_{ij}^l$ as a function of $\Lambda_2$ take the form
\begin{subequations}
\begin{align}
\dpart{W_{12}^l}{\Lambda_2} &=  \frac{m_1n_2}{m_0} \left(1+\alpha_{12}\dpart{}{\alpha_{12}}\right) \lapc{1/2}{l}{\alpha_{12}},\label{eq:dW12l}\\
\dpart{W_{23}^l}{\Lambda_2} &= - \frac{m_3n_2\alpha_{23}^2}{m_0} \dpart{\lapc{1/2}{l}{\alpha_{23}}}{\alpha_{23}}\label{eq:dW23l}.
\end{align}
\end{subequations}
Similarly, the derivatives of the factors $V_{ij,\lessgtr}$ are
\begin{subequations}
	\begin{align}
		\dpart{V_{12,<}^{l}}{\Lambda_2} &= -\frac{m_1n_2}{m_0}\sqrt{\frac{2}{\Lambda_1}} \left[l+1+(l+2)\alpha_{12}\dpart{}{\alpha}+\frac{\alpha_{12}^2}{2}\dpart{^2}{\alpha^2}\right]\lapc{1/2}{l+1}{\alpha_{12}}\\
		\dpart{V_{12,>}^{l}}{\Lambda_2} &= \frac{m_1n_2}{m_0}\sqrt{\frac{2}{\Lambda_2}} \left[\frac{10l+1}{8}+\left(l+\frac{13}{8}\right)\alpha_{12}\dpart{}{\alpha}+\frac{\alpha_{12}^2}{2}\dpart{^2}{\alpha^2}\right]\lapc{1/2}{l}{\alpha_{12}}\\
		\dpart{V_{23,<}^{l}}{\Lambda_2} &= \frac{m_3n_2\alpha_{23}}{m_0}\sqrt{\frac{2}{\Lambda_2}} \left[\frac{l+1}{4}+\right(l+\frac{11}{8}\left)\alpha_{23}\dpart{}{\alpha}+\frac{\alpha_{23}^2}{2}\dpart{^2}{\alpha^2}\right]\lapc{1/2}{l+1}{\alpha_{23}}\hspace{-.3cm}\\
		\dpart{V_{23,>}^{l}}{\Lambda_2} &= -\frac{m_3n_2\alpha_{23}}{m_0}\sqrt{\frac{2}{\Lambda_3}} \left[(l+1)\alpha_{23}\dpart{}{\alpha}+\frac{\alpha_{23}^2}{2}\dpart{^2}{\alpha^2}\right]\lapc{1/2}{l}{\alpha_{23}}\label{app:dV23ext}
	\end{align}
\end{subequations}

A few derivatives of \(S_{ij}\) and \(U_{ij}\) (see Appendix \ref{app:2ndorder}) are useful in this study. I regroup them here for convenience
\begin{subequations}
	\begin{align}
	\dpart{S_{12}^l}{x_2} &= -\frac{m_1n_2}{m_0}\left(f_2^l \bar{x}_2 +\sqrt{\frac{\Lambda_2}{\Lambda_1}}f_{10}^l\bar{x_1}\right)\\
	\dpart{S_{23}^l}{x_2} &= -\frac{m_3n_2\alpha_{23}}{m_0}\left(f_2^l \bar{x}_2 +\sqrt{\frac{\Lambda_2}{\Lambda_3}}f_{10}^l\bar{x_3}\right)\\
	\dpart{\bar{U}_{12}^l}{\bar{x}_2} &= -\frac{m_1n_2}{m_0}\left(2f_{53}^l \bar{x}_2 +\sqrt{\frac{\Lambda_2}{\Lambda_1}}f_{49}^l\bar{x_1}\right)\\
	\dpart{\bar{U}_{23}^l}{\bar{x}_2} &= -\frac{m_3n_2\alpha_{23}}{m_0}\left(2f_{45}^l \bar{x}_2 +\sqrt{\frac{\Lambda_2}{\Lambda_3}}f_{49}^l\bar{x_3}\right)
	\end{align}
	\end{subequations}

\section{Expression the perturbation part and of the functions $g_{j\lessgtr}^{k_1,k_3}$}
\label{app:gfun}
After developing the terms for Eqs. \eqref{eq:termsPoissonbracket} and \eqref{eq:termsPoissonbracket2}, the expression of the perturbation part of the Hamiltonian takes the form
\begin{align}
	\epsilon^2\H_{2,\mathrm{res}} = \frac{1}{n_2}
	&\left(\frac{W_{12}^{k_1}}{\perrat{12}^{-1}-1}\dpart{V_{23,<}^{k_3-1}}{\Lambda_2}x'_2 +\frac{W_{12}^{k_1}}{\perrat{12}^{-1}-1}\dpart{V_{23,>}^{k_3-1}}{\Lambda_2}x'_3 \right.& \nnb
	&\left.- \frac{k_3-1}{k_1(\perrat{12}^{-1}-1)}\dpart{W_{12}^{k_1}}{\Lambda_2}V_{23,<}^{k_3-1}x'_2 - \frac{k_3-1}{k_1(\perrat{12}^{-1}-1)}\dpart{W_{12}^{k_1}}{\Lambda_2}V_{23,>}^{k_3-1}x'_3 \right.& \nnb
	&\left. -\frac{W_{23}^{k_3}}{1-\perrat{23}}\dpart{V_{12,<}^{-k_1}}{\Lambda_2}x'_1  -\frac{W_{23}^{k_3}}{1-\perrat{23}}\dpart{V_{12,>}^{-k_1}}{\Lambda_2}x'_2\right.&\nnb
	&\left. + \frac{k_1-1}{k_3(1-\perrat{23})}\dpart{W_{23}^{k_3}}{\Lambda_2}V_{12,<}^{-k_1}x'_1 + \frac{k_1-1}{k_3(1-\perrat{23})}\dpart{W_{23}^{k_3}}{\Lambda_2}V_{12,>}^{-k_1}x'_2\right.&\nnb
	&\left. +\frac{3(k_3-1)W_{12}^{k_1}V_{23,<}^{k_3-1}x'_2}{k_1n_2\Lambda_2(\perrat{12}^{-1}-1)^2}+ \frac{3(k_3-1)W_{12}^{k_1}V_{23,>}^{k_3-1}x'_3}{k_1n_2\Lambda_2(\perrat{12}^{-1}-1)^2}	\right.&\nnb
	&\left. +\frac{3(k_1-1)W_{23}^{k_3}V_{12,<}^{-k_1}x'_1}{k_3n_2\Lambda_2(1-\perrat{23})^2}+ \frac{3(k_1-1)W_{23}^{k_3}V_{12,>}^{-k_1}x'_2}{k_3n_2\Lambda_2(1-\perrat{23})^2}	\right.\nnb
	&\left. -\frac{V_{12,>}^{-k_1}}{k_3(1-\perrat{{23}})}\dpart{S_{23}^{-k_3}}{x_2}+\frac{V_{12,>}^{k_1}}{k_1\perrat{12}^{-1}-(k_1+1)}\dpart{\bar{U}_{23}^{k_3-2}}{\bar{x}_2}\right.\nnb
	&\left. +\frac{V_{23,<}^{k_3-1}}{k_1(\perrat{12}^{-1}-1)}\dpart{S_{12}^{k_1}}{x_2}-\frac{V_{23,<}^{-(k_3+1)}}{k_3+1-k_3\perrat{23}}\dpart{\bar{U}_{12}^{-k_1}}{\bar{x}_2}\right) + c.c..
	\label{eq:epsH2full}
\end{align}

From Eq. \eqref{eq:epsH2full} as well as the expression of the derivatives from Appendix \ref{app:derivatives}, we have

\begin{subequations}
\begin{align}
	g_{1<}^{k_1,k_3} =&  \frac{1}{1-\perrat{23}}\left[ \left(-k_1+1+(-k_1+2)\alpha_{12}\dpart{}{\alpha}+\frac{\alpha_{12}^2}{2}\dpart{^2}{\alpha^2}\right)\lapc{1/2}{-k_1+1}{\alpha_{12}}\lapc{1/2}{k_3}{\alpha_{23}} \right.\\
	&\left. + \frac{k_1-1}{k_3}\left(\frac{3}{2(1-\perrat{23})}+\alpha_{23}\dpart{}{\alpha}\right)\lapc{1/2}{k_3}{\alpha_{23}} \left(-k_1+1+\frac{\alpha_{12}}{2}\dpart{}{\alpha}\right)\lapc{1/2}{-k_1+1}{\alpha_{12}}\right]\nnb
	&+\frac{f_{10}^{k_1}}{k_1(\perrat{12}^{-1}-1)}\left(k_3+\frac{\alpha_{23}}{2}\dpart{}{\alpha}\right)\lapc{1/2}{k_3}{\alpha_{23}}-\frac{f_{49}^{-k_1}}{k_3+1-k_3\perrat{23}}\left(-k_3+\frac{\alpha_{23}}{2}\dpart{}{\alpha}\right)\lapc{1/2}{-k_3}{\alpha_{23}}\nnb
	g_{1>}^{k_1,k_3} =& \frac{1}{1-\perrat{23}}\left[\left(\frac{1-10k_1}{8}+(-k_1+\frac{13}{8})\alpha_{12}\dpart{}{\alpha}+\frac{\alpha_{12}^2}{2}\dpart{^2}{\alpha^2}\right)\lapc{1/2}{-k_1}{\alpha_{12}}\lapc{1/2}{k_3}{\alpha_{23}}\right.\hspace{-0.3cm}\\
	&\left. +\frac{k_1-1}{k_3}\left(\frac{3}{2(1-\perrat{23})}+\alpha_{23}\dpart{}{\alpha}\right)\lapc{1/2}{k_3}{\alpha_{23}} \left(-k_1+\frac{1}{2}+\frac{\alpha_{12}}{2}\dpart{}{\alpha}\right)\lapc{1/2}{-k_1}{\alpha_{12}}\right]\nnb
	&-\frac{f_{2}^{k_1}}{k_1(\perrat{12}^{-1}-1)}\left(k_3+\frac{\alpha_{23}}{2}\dpart{}{\alpha}\right)\lapc{1/2}{k_3}{\alpha_{23}}+\frac{2f_{53}^{-k_1}}{k_3+1-k_3\perrat{23}}\left(-k_3+\frac{\alpha_{23}}{2}\dpart{}{\alpha}\right)\lapc{1/2}{-k_3}{\alpha_{23}}\nonumber
\end{align}
\end{subequations}
	\begin{subequations}
		\begin{align}
	g_{3<}^{k_1,k_3} =& \frac{1}{\perrat{12}^{-1}-1}\left[ \left(\frac{k_3-3}{4}+(k_3+\frac{3}{8})\alpha_{23}\dpart{}{\alpha}+\frac{\alpha_{23}^2}{2}\dpart{^2}{\alpha^2}\right)\lapc{1/2}{k_3}{\alpha_{23}}\lapc{1/2}{k_1}{\alpha_{12}}+\right.\\
	&\left. \frac{k_3-1}{k_1}\left(\frac{3}{2(\perrat{12}^{-1}-1)}+\alpha_{12}\dpart{}{\alpha}\right)\lapc{1/2}{k_1}{\alpha_{12}} \left(k_3+\frac{\alpha_{23}}{2}\dpart{}{\alpha}\right)\lapc{1/2}{k_3}{\alpha_{23}}\right]\nnb
	&+\frac{f_{2}^{-k_3}}{k_3(1-\perrat{23})}\left(-k_1+\frac{1}{2}+\frac{\alpha_{12}}{2}\dpart{}{\alpha}\right)\lapc{1/2}{k_1}{\alpha_{12}}-\frac{2f_{45}^{k_3-2}}{k_1\perrat{12}^{-1}-k_1-1}\left(k_1+\frac{1}{2}+\frac{\alpha_{12}}{2}\dpart{}{\alpha}\right)\lapc{1/2}{k_1}{\alpha_{12}},\nnb
	g_{3>}^{k_1,k_3} =&\frac{1}{\perrat{12}^{-1}-1}\left[\left(k_3\alpha_{23}\dpart{}{\alpha}+\frac{\alpha_{23}^2}{2}\dpart{^2}{\alpha^2}\right)\lapc{1/2}{k_3}{\alpha_{23}}\lapc{1/2}{k_1}{\alpha_{12}}+\right.\\
	&\left. \frac{k_3-1}{k_1}\left(\frac{3}{2(\perrat{12}^{-1}-1)}+\alpha_{12}\dpart{}{\alpha}\right)\lapc{1/2}{k_1}{\alpha_{12}} \left(k_3-\frac{1}{2}+\frac{\alpha_{23}}{2}\dpart{}{\alpha}\right)\lapc{1/2}{k_3}{\alpha_{23}}\right]\nnb
	&-\frac{f_{2}^{-k_3}}{k_3(1-\perrat{23})}\left(-k_1+\frac{1}{2}+\frac{\alpha_{12}}{2}\dpart{}{\alpha}\right)\lapc{1/2}{k_1}{\alpha_{12}}+\frac{f_{49}^{k_3-2}}{k_1\perrat{12}^{-1}-k_1-1}\left(k_1+\frac{1}{2}+\frac{\alpha_{12}}{2}\dpart{}{\alpha}\right)\lapc{1/2}{k_1}{\alpha_{12}}\nonumber.
\end{align}
\end{subequations}

\bibliographystyle{spbasic}      %
\bibliography{1st3plMMR.bib}   %

\end{document}